

\input amstex

\documentstyle{amsppt}

\NoRunningHeads
\TagsOnRight

\topmatter

\title
  Quasilocal Formalism for Dilaton Gravity\\
  with Yang-Mills Fields
\endtitle
\author
  Jolien D. E. Creighton and Robert B. Mann
\endauthor
\address
  Department of Applied Mathematics and Theoretical Physics,
  Silver Street, Cambridge, England.  CB3 9EW
\endaddress
\curraddr
  Department of Physics, University of Waterloo,
  Waterloo, Ontario.
\endcurraddr
\email
  jolien\@avatar.uwaterloo.ca, rbm20\@amtp.cam.ac.uk
\endemail
\date
  25 July, 1995
\enddate
\thanks
  This work was supported by NSERC, Canada.
\endthanks
\subjclass
  Primary 83C40; Secondary 83C57, 81T13
\endsubjclass
\abstract
  We present a quasilocal formalism,
  based on the one proposed by Brown and York,
  for dilaton gravity with Yang-Mills fields.
  For solutions possessing sufficient symmetry,
  we define conserved quantities such as mass,
  angular momentum, and charge.  We also present
  a micro-canonical action and use it to arrive
  at a quasilocal version of the first law of
  thermodynamics for static systems containing
  a black hole.
\endabstract

\endtopmatter

\document

\head 1\enspace Introduction\endhead
Suppose a gravitating system\footnote{In
this paper, we shall assume a four dimensional
spacetime for definiteness.} is surrounded
by some spatial surface.  On this surface,
it is possible to define quantities that will
yield information about the system contained.
These quantities we call quasilocal quantities.
In the past it has been the practice to determine
analogous quantities in an asymptotic limit.
However, this procedure must be adapted for
the specific asymptotic behaviour of the spacetime
under consideration; it is unclear how to proceed
except when the spacetime is asymptotically flat.
Furthermore, when we study the thermodynamics
of a gravitating system, the quasilocal surface
takes on an important physical significance as
the thermodynamic wall containing the system.
Thus, we find it important to develop a formalism
for constructing quasilocal quantities on the
boundary of a gravitating system, as such
quantities can then be identified directly as
the thermodynamic variables of the system.

An excellent quasilocal formalism was developed
by Brown and York \cite{Brown and York (1993a)}
for General Relativity.  In this formalism, a
surface stress energy momentum tensor was
constructed by a functional analysis of the action
for General Relativity, and, from this tensor,
a quasilocal energy can
be found.  In addition, for spacetimes possessing
sufficient symmetry, a conserved mass (which
differs, in general, from the quasilocal energy)
and angular momentum can be defined.  In their
following paper \cite{Brown and York (1993b)},
they used the Feynman path integral technique of
statistical mechanics to demonstrate the connection between
canonical and thermodynamic conjugacy.  They showed
how to construct a ``microcanonical'' action
for which the ``extensive'' variables are fixed
on the boundary of the system.  From this form
of the action, the entropy and the first law of
thermodynamics (at the zeroth order of quantum
corrections) follow easily.

Our purpose is to extend this formalism to gravity
theories involving a dilaton.  In addition, we
will include a Yang-Mills field in our analysis
to demonstrate the r\^oles of matter in the formalism.
This paper is a summary of the formalism; a more
complete discussion is given in our recent paper
\cite{Creighton and Mann (1995)}.

\head 2\enspace The Quasilocal Formalism\endhead
Assume that the region of spacetime under
consideration, $\Cal M$ (the ``gravitating
system''), is topologically the direct product
of a spacelike hypersurface, $\varSigma$,
with a real interval.  The boundary
$\partial{\varSigma}$ is the quasilocal surface
that surrounds the system, and $\Cal T$ is the
history of this surface.  We take our dynamical
fields to be the spacetime metric, $g_{\mu\nu}$,
a scalar field we call the dilaton, $\varPhi$,
and a Yang-Mills potential
${\frak{A}}_\mu{}^{\frak{a}}$.  The gauge covariant
derivative operator is
$({\frak{D}}_\mu)^{\frak{a}}{}_{\frak{b}}=\nabla_\mu
\delta^{\frak{a}}{}_{\frak{b}}+{\frak{f}}^{\frak{a}}
{}_{\frak{b}\frak{c}}{\frak{A}}_\mu{}^{\frak{c}}$
where ${\frak{f}}^{\frak{a}}{}_{\frak{b}\frak{c}}$
are the structure constants of the gauge
group.\footnote{From these structure constants, we
are able to define a Killing metric that will allow
us to raise and lower gauge group indices.}
The action we take is:
$$
  \aligned
    S &= \int_{\Cal M} d^4\!x \sqrt{-g}\,
        \bigl( D(\varPhi) R(g)
        + H(\varPhi)(\nabla\varPhi)^2 + V(\varPhi)
        - \hbox{$1\over4$} W(\varPhi)
        {\frak{F}}^{\mu\nu}{}_{\frak{a}}
        {\frak{F}}_{\mu\nu}{}^{\frak{a}} \bigr) \\
      &\qquad - 2\int_{\Cal T} d^3\!x \sqrt{-\gamma}\,
        D(\varPhi) \hbox{tr}(\varTheta)
        - 2\int_{\varSigma} d^3\!x \sqrt{h}\,
        D(\varPhi) \hbox{tr}(K).
  \endaligned
  \tag 1
$$
Here, $R(g)$ is the Ricci scalar of the metric
$g_{\mu\nu}$ and ${\frak{F}}_{\mu\nu}{}^{\frak{a}}$
is the curvature of the gauge covariant derivative
operator.  The induced metrics on $\Cal T$ and $\varSigma$
are $\gamma_{ij}$ and $h_{ij}$ respectively, while the
extrinsic curvatures of these surfaces are
$\varTheta_{ij}$ and $K_{ij}$ respectively.  The functions
$D(\varPhi)$, $H(\varPhi)$, $V(\varPhi)$, and $W(\varPhi)$
are arbitrary functions of the dilaton alone (and none
of its derivatives) that specify the theory.
\footnote{The action functional can also posess an
arbitrary functional of the boundary fields, but we shall
ignore this here.}

Under variations of the field configurations for which the boundary
configurations are fixed the induced variation in the action of
equation (1) is minimized when the equations of motion are satisfied.
Under more arbitrary variations, surface terms proportional to the
variations of the fields on the boundaries will appear along with the
equation of motion terms.  The coefficients of the boundary
variations of the fields are identified as the momenta conjugate
to those surface fields.  The phase space on $\varSigma$ is
given by $\{(p^{ij},h_{ij}),({\frak{P}}^i{}_{\frak{a}},
{\frak{A}}_i{}^{\frak{a}}),(P,\varPhi)\}$ so that, for example,
$\pi^{ij}=-(\delta S/\delta h_{ij})_{c\ell}$ (where
the ``$c\ell$'' indicates evaluation on a classical solution).
There will also be momenta conjugate to the fields on the
$\Cal T$ boundary; its phase space co\"ordinates are
$\{(\pi^{ij},\gamma_{ij}),({\frak{K}}^i{}_{\frak{a}},
{\frak{U}}_i{}^{\frak{a}}),(\varPi,\phi)\}$.  (We have written
the gauge and dilaton fields on $\Cal T$ as
${\frak{U}}_i{}^{\frak a}$ and $\phi$ respectively.)

Our primary interest will be in the $\Cal T$ boundary
which is the history of our quasilocal surface.  Parameterize
the time interval with $t$ whose corresponding affine vector
is $t^\mu=Nu^\mu+N^\mu$ where $u^\mu$ is the unit normal
to the leaves, $\varSigma_t$, of the foliation, and $N$ and
$N^\mu$ are the lapse and shift vectors.  On the quasilocal
surfaces, $\partial\varSigma_t$, there are induced metrics
$\sigma_{ij}=\gamma_{ij}+u_iu_j$.  We can decompose the fields
and conjugate momenta on $\Cal T$ into portions parallel to
the unit normals, $u^i$ (which we assume lie along $\Cal T$),
and projections onto the surfaces $\partial\varSigma$.
The metric, $\gamma_{ij}$, is thus decomposed into functions
of the lapse, the shift, and the induced metric $\sigma_{ab}$
in the usual way.  The gauge field, ${\frak{U}}_i{}^{\frak{a}}$
is decomposed into a scalar, $\frak{V}$, and a vector on
$\partial\varSigma$, ${\frak{W}}_a{}^{\frak{a}}$.  Similarly
we decompose the momenta.  Define, on $\partial\varSigma$,
${\Cal E}=\sqrt{\sigma}u_iu_j\tau^{ij}$, and,
${\Cal S}^{ab}=\sqrt{\sigma}\sigma^a_i\sigma^b_j\tau^{ij}$
as the quasilocal surface energy and stress density respectively.
Here, $\tau^{ij}=2\pi^{ij}/\sqrt{-\gamma}$ is the surface stress
energy momentum tensor on $\Cal T$.  Similarly, define
${\Cal Q}_{\frak{a}}=\sqrt{\sigma}W(\varPhi)n_i{\frak{E}}^i
{}_{\frak{a}}$ and ${\Cal I}_a{}^{\frak{a}}=\sqrt{\sigma}W(\varPhi)
\epsilon_a{}^{ij}n_i{\frak{B}}_j{}^{\frak{a}}$ as the quasilocal
Yang-Mills surface charge and current densities respectively.
Here, $n^i$ is the unit normal to $\varSigma$ and
${\frak{E}}_i{}^{\frak{a}}$ and ${\frak{B}}_i{}^{\frak{a}}$
are the usual electric and magnetic (spatial) vectors constructed
from the Yang-Mills field strength.  The surface momentum
density is given by
${\Cal J}^a=-\sqrt{\sigma}u_i\sigma^a_j\tau^{ij}
+{\Cal Q}^{\frak{a}}{\frak{W}}^a{}_{\frak{a}}$;
note that it includes an EMF term from the Yang-Mills field.
Finally, a surface dilaton density is given by
${\Cal F}=\varPi/N$.  Explicit expressions for these
quantities are given elsewhere \cite{Creighton and Mann(1995)}.

With these definitions, the variation of the action
on the boundary $\Cal T$ is
$$
  \delta S |_{\Cal T} = \int_{\Cal T} d^3\!x \bigl( -{\Cal E}
  \delta N + {\Cal J}_a \delta N^a - {\Cal Q}_{\frak{a}}
  \delta (N{\frak{V}}^{\frak{a}}) + N ( \hbox{$1\over2$}
  {\Cal S}^{ab} \delta \sigma_{ab} + {\Cal I}^a{}_{\frak{a}}
  \delta {\frak{W}}_a{}^{\frak{a}} + {\Cal F} \delta \phi ) \bigr).
  \tag 2
$$
Any variable that is a function of the phase space co\"ordinates
(on $\varSigma$) only is called an {\it extensive\/} variable.
The variables $\Cal E$, ${\Cal J}_a$, ${\Cal Q}_{\frak{a}}$,
$\sigma_{ab}$, ${\frak{W}}_a{}^{\frak{a}}$, and $\phi$ are all
extensive.  Conversely, a variable that is not a function on
the phase space of $\varSigma$ is known as an {\it intensive\/}
variable.  Functions of the lapse and the shift are examples of
intesive variables.  We see in equation (2) that the first three
terms of the integrand involve variations of intensive variables
while the last three terms involve variations of extensive variables.
We shall return to this important point in the next section.

The quasilocal surface energy, momentum, and charge densities
yield valuable information about the enclosed spacetime when
it possesses sufficient symmetries.  If the boundary, $\Cal T$,
admits a timelike, surface forming, Killing vector (such that
the Lie derivative of all the fields along this vector vanish),
and if the matter stress energy tensor is negligible on the
quasilocal boundary, then the quantity ${\Bbb{M}}=\oint d^2\!x
N{\Cal E}$ (where $\oint$ represents an integral over the quasilocal
surface, $\partial\varSigma$) is a conserved quantity representing
the mass contained within the boundary.  By ``conserved'' we mean
that the mass is independent of the foliation of $\Cal T$.
Similarly, a conserved angular momentum can be associated with the
interior of the boundary under similar conditions, but where
the Killing vector, $\varphi^a$, must now be spacelike and azimuthal.
If the quasilocal boundary contains the orbits of this vector, then
the angular momentum is ${\Bbb J}=\oint d^2\!x {\Cal J}_a\varphi^a$.
Finally, a conserved Yang-Mills charge can be associated with
each Lie-algebra-valued scalar field, ${\frak{k}}^{\frak{a}}$, that
is covariantly constant: $({\frak{D}}_\mu)^{\frak{a}}{}_{\frak{b}}
{\frak{k}}^{\frak{b}}=0$.  The conserved charge is then
${\Bbb Q}[{\frak{k}}]=\oint d^2\!x {\Cal Q}_{\frak{a}}
{\frak{k}}^{\frak{a}}$.  This is the analog of Gauss' law.
By {\it quasilocal energy}, we shall mean the quantity
${\Bbb E}=\oint d^2\!x {\Cal E}$.  Notice that this differs from
the quasilocal mass and is not necessarily a conserved quantity.
However, it will be interpreted at the thermodynamic internal
energy of the system.

\head 3\enspace The Microcanonical Action\endhead
The action that we have considered thus far is the appropriate
action for the generation of the equations of motion when, under
variations, the fields on the initial and final spacelike
hypersurfaces are fixed as are various quantities on the quasilocal
boundary.  However, we have seen that the quantities that must
be fixed on the quasilocal boundary are a mixture of extensive and
intensive variables.  The {\it microcanonical action\/} will be
that action for which only extensive variables need be fixed on
the quasilocal boundary under variations.  It differs from the
original action by a boundary term:
$$
  S_{\roman{m}} = S + \int_{\Cal T} d^3\!x ( {\Cal E} N
  - {\Cal J}_a N^a + {\Cal Q}_{\frak{a}} N {\frak{V}}^{\frak{a}} )
  \tag 3
$$
so that
$$
  \delta S_{\roman{m}}|_{\Cal T} = \int_{\Cal T} d^3\!x N (
  \delta {\Cal E} - \omega^a \delta {\Cal J}_a + {\frak{V}}^{\frak{a}}
  \delta {\Cal Q}_{\frak{a}} + \hbox{$1\over2$} {\Cal S}^{ab}
  \delta \sigma_{ab} + {\Cal I}^a{}_{\frak{a}} \delta
  {\frak{W}}_a{}^{\frak{a}} + {\Cal F} \delta \phi )
  \tag 4
$$
where $\omega^a=N^a/N$ is interpreted as the angular velocity of
observers with zero vorticity on $\Cal T$.  The covariant form
of the microcanonical action is just the same as equation (1), but
now the $\Cal T$ boundary term is:
$$
  S_{\roman{m}}|_{\Cal T} = \int_{\Cal T} d^3\!x \sqrt{-\gamma}\,
  \bigl( D(\varPhi) t_\mu \varTheta^{\mu\nu} \partial_\nu t -
  2 n^\mu\partial_\mu D(\varPhi) - W(\varPhi)(n_\mu{\frak{F}}^{\mu\nu}
  {}_{\frak{a}}\partial_\nu t)(t^\mu{\frak{A}}_\mu{}^{\frak{a}})
  \bigr).
  \tag 5
$$

We now use the microcanonical action to find an expression for
the entropy of systems containing a stationary\footnote{Stationarity
of all the fields is a necessary condition of thermodynamic
equilibrium.} black hole.  Since we will use path integral methods,
it will be useful to adopt a ``Euclidean'' notation.  In the
Euclideanization prescription, we shall construct a new microcanonical
action functional, $I_{\roman{m}}=-{\roman{i}}S_{\roman{m}}$,
so that the phase of the path integral is $\exp(-I_{\roman{m}})$.
The imaginary factor is absorbed into the intensive variables
(specifically, the lapse) so that the extensive variables are
invariant under the Wick rotation.

The quantum mechanical microcanonical density matrix is defined as
the path integral over all fields in $\Cal M$ with the phase weighting
$\exp(-I_{\roman{m}})$.  It is a functional of the variables
held fixed on the initial and final spatial hypersurfaces,
$\varSigma_{\roman{initial}}$ and $\varSigma_{\roman{final}}$,
as well as the extensive variables held fixed on $\Cal T$.
The most important statistic of this density matrix is the
partition function which, in the microcanonical ensemble, is known
as the density of states.  This is obtained by identifying the
initial and the final spatial hypersurfaces (with some period,
$\Delta t$) and ``tracing over'' the field configurations on these
identified surfaces.  The density of states, then, is a functional
of the extensive variables on the qusilocal surface.

Because the system contains a black hole, the foliation becomes
degenerate on the event horizon.  When the initial and final spatial
hypersurfaces have been identified, the complex manifold has the
topology of a cone$\times S^2$ (we assume that it is posible to
foliate from the event horizon to the quasilocal boundary with
parameter~$r$).  The conical singularity at the event horizon is
removed by fixing the period of identification to the value
$\Delta t=2\pi/\kappa_{\scriptscriptstyle\roman{H}}$ where
$\kappa_{\scriptscriptstyle\roman{H}}^2=h^{ij}(\partial_i N)
(\partial_j N)$ (evaluated on the event horizon) is the surface
gravity of the event horizon.

We now find the entropy of the gravitating system to ``zeroth
order'' in quantum corrections.  In this approximation, the density
of states is just the path integral phase evaluated on a classical
solution, so the entropy is
${\Bbb{S}}\approx-I_{\roman{m}}|_{c\ell}$.  We thus need to evaluate
the Euclideanized microcanonical action at its (complex) extremal
value.  The event horizon is not included in the system as it acts
as a one-way membrane for observers orbiting on $\Cal T$.
This is accoplished most easily by performing a canonical
decomposition of the microcanonical action with the surface term
given in equation (5).  No boundary term is appended to the action
on the event horizon.  We find that:
$$
  \aligned
    I_{\roman{m}} &= {\roman i}\int dt\, \biggl(
    \int_\varSigma d^3\!x ( -p^{ij}{\dot{h}}_{ij}
    - {\frak{P}}^i{}_{\frak{a}} {\dot{\frak{A}}}_i{}^{\frak{a}}
    - P{\dot{\varPhi}} + {\Cal H}N + {\Cal H}_i N^i
    + {\Cal G}_{\frak{a}} {\frak{A}}_t{}^{\frak{a}} ) \\
    &\quad + \int_H d^2\!x \Bigl( 2\sqrt{\sigma}\,\bigl(
    D(\varPhi) n^i\partial_i N - N n^i\partial_i D(\varPhi) \bigr)
    - N^a {\Cal J}_a + N {\Cal Q}_{\frak{a}} {\frak{V}}^{\frak{a}}
    \Bigr) \biggr)
  \endaligned
  \tag 6
$$
where the over-dot indicates a Lie derivative with respect to time
and $\Cal H$, ${\Cal H}_i$, and ${\Cal G}_{\frak{a}}$ are the
Hamiltonian, momentum, and Gauss constraints respectively.  Since
the solution is stationary, all the time derivatives vanish and
since the action is to be evaluated on a classical solution, the
constraints also vanish.  All that is left, then is the integral
over the event horizon.  Here, we have $N=0$, $N^a=0$, and, from
the regularity conditions, $\int dt\,n^i\partial_i N=2\pi{\roman{i}}$.
Thus, we find that the entropy is
${\Bbb S}=\int d^2\!x \sqrt{\sigma}\, D(\varPhi)$
where the integral is over the event horizon.  Notice that, in the
case of General Relativity, we have $D(\varPhi)=(16\pi)^{-1}$, and
the entropy is the usual value of one quarter of the event horizon
area (in units of the rationalized Planck's constant and Newton's
constant).

The first law of thermodynamics follows immediately from
equation (4).  Define $\beta={\roman{i}}\int dt\, N$ evaluated
on the quasilocal boundary, $\partial\varSigma$.  It is the
reciprocal temperature of the system.  Notice that it is not
necessarily constant over the quasilocal surface unless the
quasilocal surface is so chosen.  The first law of thermodynamics
for the quasilocal system is:
$$
  \delta{\Bbb S} = \int_{\partial\varSigma} d^2\!x \beta (
  \delta {\Cal E} - \omega^a \delta {\Cal J}_a + {\frak{V}}^{\frak{a}}
  \delta {\Cal Q}_{\frak{a}} + \hbox{$1\over2$}{\Cal S}^{ab}
  \delta \sigma_{ab} + {\Cal I}^a{}_{\frak{a}} \delta {\frak{W}}_a
  {}^{\frak{a}} + {\Cal F} \delta \phi ).
  \tag 7
$$
When the quasilocal surface is chosen to be an isotherm, the
first term of the integral becomes the usual ``$\beta\delta{\Bbb E}$''
term.  However, since the angular velocity, for example, will not
necessarily be constant over an isotherm, it is necessary to write
the first law in integral form.

\refstyle{B}


\Refs

\ref
\by      Brown, J. D. and York, J. W.
\paper   Quasilocal energy and conserved charges derived from
         the gravitational action
\jour    Phys. Rev. D.
\vol     47
\yr      1993a
\pages   1407--1419
\endref

\ref
\by      Brown, J. D. and York, J. W.
\paper   Microcanonical functional integral for the
         gravitational field
\jour    Phys. Rev. D.
\vol     47
\yr      1993b
\pages   1420--1431
\endref

\ref
\by      Creighton, J. D. E. and Mann, R. B.
\paper   Quasilocal Thermodynamics of Dilaton Gravity
         Coupled to Gauge Fields
\jour    Preprint gr-qc/9505007
\yr      1995
\endref

\endRefs
\enddocument